\documentclass{article}

    \PassOptionsToPackage{numbers, compress}{natbib}

\usepackage[preprint]{neurips_2026}

\usepackage[utf8]{inputenc} 
\usepackage[T1]{fontenc}    
\usepackage{hyperref}       
\usepackage{url}            
\usepackage{booktabs}       
\usepackage{amsfonts}       
\usepackage{nicefrac}       
\usepackage{microtype}      
\usepackage{xcolor}         
\usepackage{enumerate}
\usepackage{amsmath}
\usepackage{graphicx}
\usepackage{enumitem}
\usepackage{tabularx}
\usepackage{array}

\title{DegradeQuery: Counterfactual Tuple Pretraining for Context-Aware PROTAC Degradation Prediction}

\author{%
  \textbf{Dong Xu\textsuperscript{1,2},
  Zhangfan Yang\textsuperscript{3},
  Jiantao Wu\textsuperscript{1},
  Zexuan Zhu\textsuperscript{1},
  Jianqiang Li\textsuperscript{1},
  Junkai Ji\textsuperscript{1,2,\textdagger}}\\[2pt]
  \small \textsuperscript{1}School of Artificial Intelligence, Shenzhen University
  \quad \textsuperscript{2}EasternDawn\\
  \small \textsuperscript{3}School of Computer Science, University of Nottingham Ningbo
}

\begin{document}

\maketitle

\begingroup
\renewcommand{\thefootnote}{}
\footnotetext{This work is supported by the Intelligent Computing Center of Shenzhen University.}
\endgroup
\begingroup
\renewcommand{\thefootnote}{\textdagger}
\footnotetext{Corresponding author: Junkai Ji (e-mail: \texttt{jijunkai@szu.edu.cn}).}
\endgroup

\begin{abstract}
Proteolysis-targeting chimeras (PROTACs) induce protein degradation by recruiting a target protein to an E3 ubiquitin ligase, making degradation a joint outcome of the degrader molecule and its biological context. Although public databases contain thousands of structured molecule--target--E3 records, degradation measurements are available for only a small fraction of them. Existing supervised approaches therefore leave most recorded chemical--biological relationships unused. We introduce DegradeQuery, a context-aware prediction framework that converts these label-missing records into a pretraining signal. Its counterfactual tuple pretraining objective contrasts recorded tuples with alternatives formed by replacing the target, the E3 ligase, or both, enabling the model to learn contextual associations without assigning activity pseudo-labels. The resulting representation is then fine-tuned to predict degradation from the complete molecule--target--E3 context. On the official PROTAC-8K benchmark, DegradeQuery achieves an area under the receiver operating characteristic curve of 0.9065 and an accuracy of 0.8500, outperforming the compared methods. Controlled analyses further show that the improvement is primarily attributable to tuple-level pretraining, can be recovered using only label-missing records, and remains complementary to protein language model representations. These findings demonstrate that incompletely labeled PROTAC databases contain useful relational supervision and provide a practical route for learning context-aware degradation predictors from scarce experimental labels.
\end{abstract}

\section{Introduction}

Proteolysis-targeting chimeras (PROTACs) have fundamentally changed how small molecules can modulate protein function.
Unlike conventional inhibitors, which typically act by occluding a binding site, PROTACs recruit a target protein into proximity with an E3 ubiquitin ligase, thereby promoting target ubiquitination and subsequent proteasomal degradation \citep{sakamoto2001protacs,lai2017induced,burslem2020proteolysis,bekes2022protac}.
This induced-proximity mechanism enables catalytic removal of the target in cells \citep{bondeson2015catalytic} and has been successfully exploited through E3 recruiters such as cereblon ligands \citep{winter2015phthalimide,lu2015hijacking}.
It further broadens the druggable target space to proteins that are difficult to modulate via classical occupancy-driven inhibition \citep{schapira2019targeted,bekes2022protac}.
At the same time, this conditional mechanism renders degradation prediction considerably more challenging.
A degrader molecule should not be modeled as having a single context-independent degradation label;
its measured effect depends jointly on the molecular structure of the degrader, the target protein, the recruited E3 ligase, and the cellular context in which the assay is performed.
PROTAC degradation is therefore not solely a molecular property but rather a context-dependent outcome of the molecule--target--E3 system.

This distinction carries direct implications for data-driven degradation modeling.
Public PROTAC resources now provide structured records comprising degrader molecules, target proteins, E3 ligases, warheads, linkers, and assay annotations \citep{weng2021protac,weng2023protac,cai2025protac,ge2025protac}.
Related resources for molecular glues similarly catalog chemically induced degradation entries by small molecule, recruited protein, and target context \citep{wang2026molgluedb,zhu2026mgtbind}.
Yet reliable degradation labels remain scarce.
Quantities such as the half-maximal degradation concentration ($DC_{50}$), maximum degradation ($D_{\mathrm{max}}$), and measured percent degradation are costly to obtain and consequently unavailable for many entries \citep{weng2021protac,weng2023protac}.
As a result, a substantial portion of available data contains chemical and biological context but lacks a complete degradation label.
Treating PROTAC prediction as a standard supervised molecular classification problem leaves this context underutilized and encourages models to learn primarily from the labeled subset, even though unlabeled entries still describe which molecules were studied in which target--E3 contexts.

Recent computational methods have advanced PROTAC degradation prediction along several directions.
DeepPROTACs demonstrated that deep learning can predict targeted degradation from PROTAC-derived representations \citep{li2022deepprotacs}.
PROTAC-Degradation-Predictor extended this line of work by incorporating assay and target context into machine-learning-based prediction \citep{ribes2024modeling}.
More recent models have moved toward explicit ternary or structure-aware representations:
PROTAC-STAN employs a structure-informed deep ternary attention framework \citep{chen2025interpretable}, and
DegradeMaster adopts a semi-supervised E(3)-equivariant graph neural network \citep{liu2025accurate}.
Complementary work on ternary complex prediction and biomolecular interaction modeling reinforces this perspective, indicating that the relevant computational object is not an isolated molecule but a multi-component interaction system \citep{xue2025se,abramson2024accurate}.
Together, these studies establish that degradation prediction benefits from modeling beyond the degrader molecule alone.

A central open question remains: how should unlabeled PROTAC records be leveraged?
One natural strategy is to assign pseudo-labels to unlabeled examples and train on the expanded set, as in semi-supervised degradation prediction \citep{liu2025accurate}.
While effective in practice, this approach treats unlabeled data as though the primary missing piece were a degradation label.
Such a framing is not the only way to exploit these records.
The most reliable signal in an unlabeled PROTAC record is not a latent binary activity label but the recorded molecule--target--E3 context itself.
Even without a measured degradation value, an entry typically specifies a PROTAC molecule, a target protein, and an E3 ligase.
We interpret this tuple as observed design-context evidence: it indicates that a molecule was recorded or studied in relation to a target and an E3 recruiter, but it does not by itself imply active degradation, ternary-complex formation, or biological feasibility.

This observation motivates a different form of self-supervision.
Rather than asking a model to impute a degradation label, we ask it to distinguish observed molecule--target--E3 records from counterfactual records formed by replacing the target, the E3 ligase, or both.
The model thereby learns a tuple-conditioned pretraining signal: which recorded tuples receive higher scores than sampled alternatives under the chosen corruption process.
This idea is connected to the broader principle of learning representations from contrastive structure in unlabeled data \citep{oord2018representation,chen2020simple}.
In molecular representation learning, contrastive objectives have been applied to learn robust graph-level or molecular representations \citep{you2020graph,wang2022molecular}.
However, PROTAC degradation demands a more specific signal than molecular invariance alone.
A given molecule may serve as an effective degrader in one target--E3 context yet fail in another.
The self-supervised signal should therefore be defined at the level at which PROTAC records are specified and degradation labels are ultimately assigned: the molecule--target--E3 tuple.

We introduce \textbf{DegradeQuery}, a single-model framework built around this principle.
DegradeQuery first learns from unlabeled PROTAC records through counterfactual tuple pretraining.
Given an observed tuple $(m_i,t_i,e_i)$, where $m_i$ denotes the PROTAC molecule, $t_i$ the target protein, and $e_i$ the E3 ligase, the model scores this tuple against counterfactual alternatives, including $(m_i,t_j,e_i)$, $(m_i,t_i,e_j)$, and $(m_i,t_j,e_k)$, in which one or both biological components are replaced.
This objective trains the representation on molecule--target--E3 records before any degradation label is introduced.
The resulting pretraining score is a tuple plausibility score relative to sampled counterfactuals, not a degradation probability.
The pretrained model is then fine-tuned on labeled degradation data for binary (active versus inactive) classification.
Notably, DegradeQuery does not generate pseudo-labels for unlabeled records and does not rely on teacher models, distillation, or ensembles.
Unlabeled data enter the learning process solely through their observed tuple structure.

This design redefines the role of unlabeled PROTAC data.
Rather than treating unlabeled records as incomplete supervised examples, DegradeQuery treats them as structured tuple observations.
The distinction matters because degradation labels are sparse and assay-dependent, whereas the tuple itself is usually available and reflects the molecule, target, and E3 components required by the induced-proximity mechanism \citep{lai2017induced,bekes2022protac}.
Generic molecular self-supervision can produce robust molecular representations, yet it cannot, on its own, encode which target and E3 contexts accompany a molecule in the observed PROTAC records.
Counterfactual tuple pretraining addresses this gap by situating the molecule within the recorded target--E3 context before supervised degradation labels are used.

We evaluate DegradeQuery on the PROTAC-8K benchmark and in additional settings that probe the source and scope of its improvement.
On the official split used by DegradeMaster, DegradeQuery achieves the highest reported AUROC and accuracy among the compared methods without relying on pseudo-labels, teacher models, distillation, or ensembles.
Matched runs show that the gain is stable on this split, persists when every labeled row is removed from pretraining, and remains present with frozen ESM2-650M protein representations.
Component ablations show that tuple pretraining contributes beyond molecule-only self-supervision, while repeated target--E3 group holdouts reveal substantial variation across held-out contexts.
Throughout the paper, tuple plausibility denotes a sampler-relative score learned from recorded tuples and counterfactual replacements; it is not evidence of ternary-complex formation, degradative activity, or biological feasibility.

This work makes three contributions:
\begin{enumerate}[
    label=\arabic*.,
    labelindent=10pt,
    leftmargin=*,
    labelsep=0.5em,
    itemsep=0pt,
    topsep=2pt
]
    \item[(a)] We formulate unlabeled PROTAC records as structured molecule--target--E3 observations rather than incomplete examples awaiting pseudo-labels.

    \item[(b)] We propose DegradeQuery, which employs counterfactual tuple pretraining to learn tuple-conditioned PROTAC representations without assigning degradation pseudo-labels, using teacher models, applying distillation, or ensembling models.

    \item[(c)] We isolate the contribution of label-missing records through five-seed paired evaluation, an unlabeled-only pretraining control, a molecule-versus-tuple component analysis, and an ESM2-650M comparison. We further characterize the variability of target--E3 group holdout performance rather than treating a single held-out split as evidence of uniformly robust extrapolation.
\end{enumerate}

\section{Related Work}

\textbf{Context-Aware PROTAC Degradation Prediction.}
Computational approaches to PROTAC degradation prediction have increasingly moved beyond molecule-only representations toward context-aware modeling. DeepPROTACs showed that targeted degradation can be predicted by jointly modeling the PROTAC molecule, the protein of interest, and the E3 ligase \citep{li2022deepprotacs}. PROTAC-Degradation-Predictor further incorporated E3, target, and assay context into degradation prediction \citep{ribes2024modeling}. More recent models make this contextual structure more explicit: PROTAC-STAN uses a structure-informed ternary attention framework to model PROTAC--target--E3 interactions \citep{chen2025interpretable}, while DegradeMaster introduces a semi-supervised E(3)-equivariant graph neural network with memory-based pseudo-labeling for unlabeled PROTAC records \citep{liu2025accurate}. Related progress in ternary complex prediction and biomolecular interaction modeling similarly suggests that the relevant computational object is often an interaction system rather than an isolated molecule \citep{xue2025se,abramson2024accurate}. These studies motivate context-aware degradation prediction; our work differs mainly in how unlabeled records are used, treating them as observed molecule--target--E3 tuples rather than as examples requiring imputed degradation labels.

\textbf{Learning from Unlabeled Molecular Data.}
Self-supervised learning has become a standard approach for extracting representations from unlabeled data. Contrastive Predictive Coding and SimCLR learn by distinguishing related views from unrelated ones \citep{oord2018representation,chen2020simple}. In graph and molecular learning, GraphCL uses graph augmentations to learn invariant graph representations \citep{you2020graph}, and MolCLR applies contrastive learning to molecular graphs for downstream molecular property prediction \citep{wang2022molecular}. These methods show that unlabeled data can provide useful training signals without task-specific labels. For PROTACs, however, molecule-level invariance alone is limited, because the same molecule can be relevant in one target--E3 context and uninformative in another. This motivates defining self-supervision at the molecule--target--E3 tuple level rather than only over augmented views of an isolated molecule.

\begin{figure*}[!t]
\centering
\includegraphics[width=\textwidth]{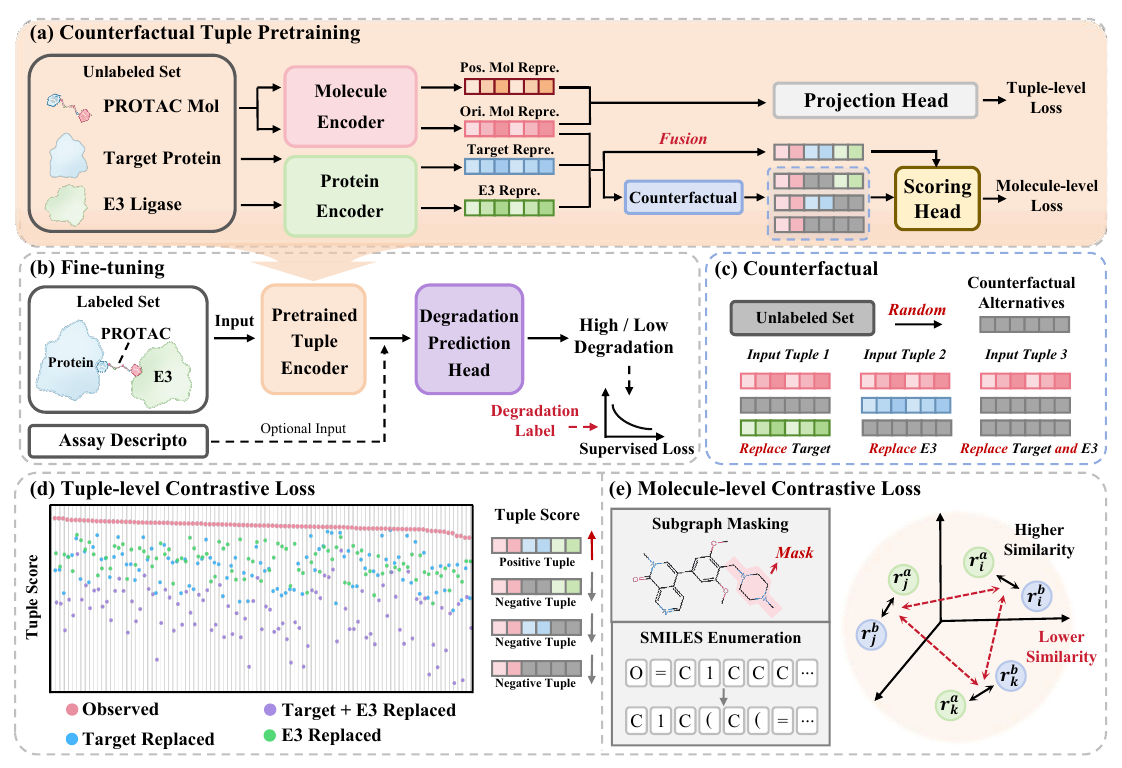}
\vspace{-6mm}
\caption{\textbf{Overview of DegradeQuery.}
DegradeQuery treats the molecule--target--E3 tuple as the unit of unlabeled supervision.
In counterfactual tuple pretraining, observed PROTAC tuples from labeled and unlabeled training records are contrasted with alternatives formed by replacing the target, the E3 ligase, or both.
This stage learns a tuple-conditioned representation without assigning pseudo-labels to unlabeled records.
The pretrained encoder is then fine-tuned on labeled degradation data for binary high/low prediction.}
\label{fig:overview}
\vspace{-4mm}
\end{figure*}

\textbf{Low-Resource Molecular Prediction.}
Low-resource learning remains a central challenge in molecular prediction, where many tasks contain only a small number of measured compounds. FS-Mol formalized this setting by aggregating activity measurements across diverse protein targets for few-shot molecular learning \citep{stanley2021fs}. Subsequent work has emphasized the role of task context: Property-Aware Relation Networks learn property-aware molecular embeddings and adaptive relation graphs \citep{wang2021property}, while context-enriched molecule representations use reference or support molecules to improve few-shot drug discovery \citep{schimunek2023context}. These studies indicate that molecular representations can depend strongly on the task or support context. In this paper, target-wise few-shot adaptation is used as an evaluation protocol for target-conditioned degradation representations, rather than as a claim of a new general-purpose meta-learning algorithm.

\section{Method}

\paragraph{Overview.}
Figure~\ref{fig:overview} summarizes the DegradeQuery workflow.
Rather than treating unlabeled PROTAC records as examples with missing degradation labels, DegradeQuery treats them as recorded molecule--target--E3 contexts.
It first learns a tuple-conditioned representation by contrasting recorded tuples with sampled target and E3 replacements, and then fine-tunes the resulting encoder on labeled high/low degradation data.
The pretraining objective is defined relative to the train-side tuple set and specified counterfactual sampler; it does not imply degradation activity, ternary-complex formation, or biological feasibility.

\subsection{Problem Formulation}

For a given data split, let $\mathcal{D}_{L}$ and $\mathcal{D}_{U}$ denote the train-side labeled and unlabeled subsets, respectively.
Each PROTAC record comprises a degrader molecule, a target protein, an E3 ligase, and, optionally, an assay descriptor such as a cell line \citep{weng2021protac,weng2023protac}.
Only a subset of records carries a degradation label \citep{liu2025accurate}.
We denote the labeled and unlabeled sets as
\begin{equation}
\mathcal{D}_{L}
=
\{(m_i,t_i,e_i,c_i,y_i)\}_{i=1}^{n_L},
\qquad
\mathcal{D}_{U}
=
\{(m_i,t_i,e_i,c_i)\}_{i=1}^{n_U},
\label{eq:data_rev}
\end{equation}
where $m_i$ is the PROTAC molecule, $t_i$ is the target protein, $e_i$ is the E3 ligase, $c_i$ is an optional assay descriptor, and $y_i\in\{0,1\}$ is the binary degradation label.

The goal is to learn a tuple-conditioned degradation predictor
\begin{equation}
p_{\theta}(y_i=1\mid m_i,t_i,e_i,c_i)
=
\sigma\!\left(g_{\theta}(m_i,t_i,e_i,c_i)\right),
\label{eq:prediction_rev}
\end{equation}
where $\sigma(\cdot)$ denotes the sigmoid function.
The unlabeled set is not treated as a collection of molecules with missing labels, but rather as a set of recorded molecule--target--E3 contexts.
We use their empirical co-occurrence structure as a self-supervised signal, while recognizing that this structure may reflect literature coverage, database construction, and medicinal-chemistry design bias rather than degradation activity.

For each data split, pretraining constructs the train-side tuple set
\begin{equation}
\mathcal{T}
=
\{(m_i,t_i,e_i):(m_i,t_i,e_i,c_i,y_i)\in \mathcal{D}_{L}\}
\cup
\{(m_i,t_i,e_i):(m_i,t_i,e_i,c_i)\in \mathcal{D}_{U}\}.
\label{eq:tuple_set_rev}
\end{equation}
Held-out test, query, or adaptation-query rows are excluded from $\mathcal{T}$ before pretraining.
Degradation labels are not used in this stage.
Both active and inactive labeled records enter $\mathcal{T}$ in the same way.
In this stage, a positive tuple means that the molecule--target--E3 context was recorded in the dataset; it does not mean that the tuple is degradation-active.

\subsection{Tuple-Conditioned Degrader Model}

DegradeQuery encodes the molecule, target protein, and E3 ligase into separate representations:
\begin{equation}
h_i^{m}=f_m(m_i),
\qquad
h_i^{t}=f_p(t_i),
\qquad
h_i^{e}=f_p(e_i).
\label{eq:encoders_rev}
\end{equation}
Here $f_m$ combines a three-layer residual message-passing network over 15-dimensional atom features with a 2,048-bit Morgan fingerprint (radius 2); the two branches are projected to 256 dimensions and summed with layer normalization.
The shared protein encoder $f_p$ maps 150 sequence features---amino-acid composition, sequence-length and unknown-residue features, and 128 hashed di-/tri-peptide counts---to 256 dimensions.
The target and E3 representations remain separate after encoding because they serve distinct roles within the tuple.

The fusion module exposes both component identities and pairwise interactions:
\begin{equation}
z_i = \phi_{\theta}\!\left(
h_i^{m} \Vert h_i^{t} \Vert h_i^{e}
\Vert h_i^{m}\!\odot h_i^{t}
\Vert h_i^{m}\!\odot h_i^{e}
\Vert h_i^{t}\!\odot h_i^{e}
\Vert |h_i^{t}-h_i^{e}|
\right),
\label{eq:tuple_fusion_rev}
\end{equation}
where $\Vert$ denotes concatenation and $\odot$ denotes elementwise multiplication.
This role-conditioned fusion is the common prediction architecture used by Sup and CTP; CTP changes the initialization through pretraining, not the downstream topology.
When the assay descriptor is available, it is incorporated via
$\tilde{z}_i
=
\psi_{\theta}(z_i,c_i)$.
The self-supervised tuple objective operates on $z_i$, whereas the downstream degradation classifier operates on $\tilde{z}_i$.
We use $c_i$ only in supervised fine-tuning because assay descriptors can be missing or inconsistently annotated for unlabeled records, whereas molecule, target, and E3 identifiers are available for tuple pretraining.

For counterfactual tuple pretraining, DegradeQuery assigns a scalar sampler-relative tuple plausibility score to each molecule--target--E3 tuple:
\begin{equation}
s_{\theta}(m_i,t_i,e_i)
=
q_{\theta}\!\left(
\phi_{\theta}\!\left(f_m(m_i),f_p(t_i),f_p(e_i)\right)
\right),
\label{eq:tuple_score_rev}
\end{equation}
where $q_{\theta}$ is a scoring head.
This score is used exclusively during pretraining.
It is meaningful only relative to the sampled counterfactual distribution and is not interpreted as a degradation probability or as an absolute biological compatibility score.

\subsection{Counterfactual Tuple Pretraining}

For each recorded tuple
$u_i^{+}=(m_i,t_i,e_i)$,
we construct counterfactual alternatives by replacing the target, the E3 ligase, or both.
Let $\mathcal{P}_{T}$ and $\mathcal{P}_{E}$ denote the empirical train-side distributions over targets and E3 ligases in the pretraining tuple set $\mathcal{T}$.
For each $u_i^{+}$, we sample $R$ replacement pairs
$\{(\tilde{t}_{i,r},\tilde{e}_{i,r})\}_{r=1}^{R}$, where
$\tilde{t}_{i,r}\sim\mathcal{P}_{T}$, $\tilde{e}_{i,r}\sim\mathcal{P}_{E}$, $\tilde{t}_{i,r}\neq t_i$, and $\tilde{e}_{i,r}\neq e_i$.
The resulting counterfactual set is
\begin{equation}
\mathcal{N}_i
=
\bigcup_{r=1}^{R}
\left\{
(m_i,\tilde{t}_{i,r},e_i),\;
(m_i,t_i,\tilde{e}_{i,r}),\;
(m_i,\tilde{t}_{i,r},\tilde{e}_{i,r})
\right\}.
\label{eq:counterfactual_set_rev}
\end{equation}
In split-defined holdout experiments, samples that would create a held-out target--E3 pair are rejected, so held-out pairs are not introduced as corrupted pretraining alternatives.

These counterfactuals are not assumed to be biologically impossible or degradation-inactive.
They are sampler-defined alternatives that may include untested but viable molecule--target--E3 combinations.
Thus, the objective is noise-contrastive rather than supervised with true negative biological examples: the recorded tuple is encouraged to score above sampled alternatives under the chosen replacement distribution.

The tuple scoring function is optimized with a contrastive ranking loss:
\begin{equation}
\mathcal{L}_{\mathrm{tuple}}
=
-\frac{1}{|\mathcal{B}|}
\sum_{i\in\mathcal{B}}
\log
\frac{
\exp\!\left(s_{\theta}(u_i^{+})/\tau\right)
}{
\exp\!\left(s_{\theta}(u_i^{+})/\tau\right)
+
\sum_{u^{-}\in\mathcal{N}_i}
\exp\!\left(s_{\theta}(u^{-})/\tau\right)
},
\label{eq:tuple_loss_rev}
\end{equation}
where $\mathcal{B}$ is a minibatch and $\tau$ is a temperature parameter.
This loss relies solely on train-side recorded tuples in $\mathcal{T}$ and does not require, infer, or assign degradation labels.

In addition, following graph and molecular contrastive learning, we employ a lightweight molecular consistency objective to encourage augmentation-invariant molecular representations \citep{you2020graph,wang2022molecular}.
Given two stochastic augmentations $a(m_i)$ and $b(m_i)$ of the same molecule, such as random subgraph masking or simplified molecular-input line-entry system (SMILES) enumeration, let
\begin{equation}
r_i^{a}=\rho_{\theta}(f_m(a(m_i))),
\qquad
r_i^{b}=\rho_{\theta}(f_m(b(m_i))),
\label{eq:molecule_views_rev}
\end{equation}
where $\rho_{\theta}$ is a projection head.
The molecule-level objective is the corresponding normalized temperature-scaled contrastive objective \citep{chen2020simple}:
\begin{equation}
\mathcal{L}_{\mathrm{mol}}
=
-\frac{1}{|\mathcal{B}|}
\sum_{i\in\mathcal{B}}
\log
\frac{
\exp\!\left(\mathrm{sim}(r_i^{a},r_i^{b})/\tau_m\right)
}{
\sum_{j\in\mathcal{B}}
\exp\!\left(\mathrm{sim}(r_i^{a},r_j^{b})/\tau_m\right)
},
\label{eq:mol_loss_rev}
\end{equation}
where $\mathrm{sim}(\cdot,\cdot)$ denotes cosine similarity and $\tau_m$ is a temperature parameter.

The full pretraining objective combines both terms:
\begin{equation}
\mathcal{L}_{\mathrm{pre}}
=
\mathcal{L}_{\mathrm{tuple}}
+
\lambda_{\mathrm{mol}}\mathcal{L}_{\mathrm{mol}}.
\label{eq:pretrain_loss_rev}
\end{equation}
The tuple loss is the primary training signal, while the molecule-level term serves as an auxiliary regularizer.

\subsection{Fine-Tuning and Prediction}

After pretraining, the tuple encoder is fine-tuned on the labeled degradation records.
The model predicts
\begin{equation}
\hat{p}_i
=
p_{\theta}(y_i=1\mid m_i,t_i,e_i,c_i)
=
\sigma\!\left(w^{\top}\tilde{z}_i+b\right),
\label{eq:finetune_prediction_rev}
\end{equation}
where $\tilde{z}_i$ is the descriptor-aware representation.

We optimize a binary classification loss over labeled examples only.
Let $p_i^{\ast}$ denote the predicted probability assigned to the true class:
\begin{equation}
p_i^{\ast}
=
y_i\hat{p}_i
+
(1-y_i)(1-\hat{p}_i).
\label{eq:pt_rev}
\end{equation}
The supervised objective is a class-weighted focal loss \citep{lin2017focal}:
\begin{equation}
\mathcal{L}_{\mathrm{sup}}
=
-\frac{1}{|\mathcal{D}_{L}|}
\sum_{i\in\mathcal{D}_{L}}
\alpha_{y_i}
(1-p_i^{\ast})^{\gamma}
\log p_i^{\ast}.
\label{eq:focal_loss_rev}
\end{equation}
When $\gamma=0$, this formulation reduces to weighted binary cross-entropy, so it subsumes the standard case as a special case of focal loss.

The two training stages can be summarized as
$
\theta_{\mathrm{pre}}
=
\arg\min_{\theta}
\mathcal{L}_{\mathrm{pre}}(\mathcal{T})$,
followed by
\begin{equation}
\theta^{\star}
=
\arg\min_{\theta}
\mathcal{L}_{\mathrm{sup}}(\mathcal{D}_{L}),
\qquad
\theta \text{ initialized from } \theta_{\mathrm{pre}}.
\label{eq:finetune_stage_rev}
\end{equation}
Unlike pseudo-label-based semi-supervised PROTAC degradation prediction \citep{liu2025accurate}, unlabeled records contribute exclusively through the tuple pretraining objective: they are used as recorded contexts, never as degradation-labeled examples.
At inference time, DegradeQuery applies the fine-tuned model directly:
$\hat{y}_i
=
\mathbb{I}\!\left[\hat{p}_i \geq \delta\right]$,
where $\mathbb{I}[\cdot]$ is the indicator function and $\delta$ is a fixed decision threshold specified by the evaluation protocol.
No teacher model, pseudo-labeling step, distillation procedure, or model ensemble is used.

\section{Experiments}
\label{sec:experiments}

\noindent We evaluate whether counterfactual tuple pretraining (CTP) improves PROTAC-8K without pseudo-labeling, whether label-missing records alone account for the gain, whether CTP remains useful with a strong protein representation, and how performance changes under target--E3 group holdout, scaffold holdout, and target-wise few-shot adaptation.

\begin{figure*}[t]
\centering
\includegraphics[width=\textwidth]{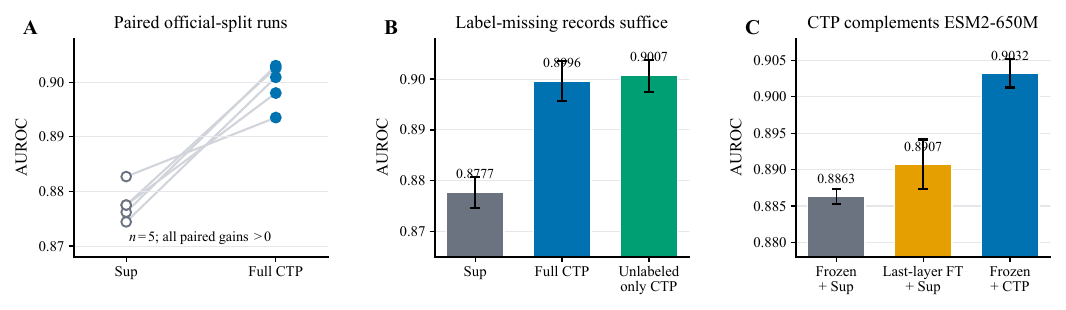}
\vspace{-5mm}
\caption{Evidence isolating the contribution of counterfactual tuple pretraining.
(A) Seed-level AUROC for the same supervised architecture without and with CTP on the official split; all five paired gains are positive.
(B) Mean $\pm$ standard deviation over five seeds. Unlabeled-only CTP removes every labeled row from pretraining and uses 7,134 label-missing records.
(C) Mean $\pm$ standard deviation over three seeds when ESM2-650M supplies the target and E3 representations inside the same role-conditioned PROTAC tuple predictor.}
\label{fig:protac8k_diagnostics}
\vspace{-4mm}
\end{figure*}

\subsection{Experimental Setup}
\label{sec:experimental_setup}

We evaluate binary high/low PROTAC degradation prediction on the PROTAC-8K benchmark used in prior work~\cite{liu2025accurate}.
The normalized dataset contains 8,636 molecule--target--E3 records, of which 1,502 are labeled and 7,134 are unlabeled.
Molecule, target, and E3 identifiers are available for all records, so PROTAC-8K is label-sparse but tuple-rich enough for self-supervised tuple pretraining.
All experiments are split-specific.
Test or query rows are never used for tuple pretraining, supervised fine-tuning, model selection, or threshold tuning.
In target--E3 group and scaffold holdouts, held-out records are excluded from both tuple pretraining and supervised training.
For target--E3 group holdout, the sampler also rejects corrupted tuples whose target--E3 pair is held out, preventing held-out biological pairs from entering pretraining as negatives.
In target-wise few-shot adaptation, support rows are used only for adaptation and query rows only for evaluation; neither is used during base pretraining or base supervised training.
We compare DegradeQuery-Sup, trained only on labeled training records, with DegradeQuery-CTP, which first pretrains on train-side molecule--target--E3 tuples, including unlabeled records, and then fine-tunes on labeled records.
CTP uses no pseudo-labels, teacher models, distillation, or ensembles.
AUROC is the primary metric, accuracy is reported for benchmark compatibility, and threshold-dependent metrics use a fixed threshold of 0.5.
Each table states its statistical unit: the new official-split reliability and unlabeled-only analyses use five paired seeds, the ESM2 and component analyses use three seeds, and the repeated target--E3 analysis uses ten independently constructed group holdouts. Metrics are reported on the 0--1 scale.

\subsection{PROTAC-8K Benchmark}
\label{sec:protac8k_benchmark}

Table~\ref{tab:protac8k_benchmark} compares DegradeQuery with prior baselines under the official PROTAC-8K protocol used by DegradeMaster~\cite{liu2025accurate}.
Unlike DegradeMaster-Semi, which uses pseudo-labeling, DegradeQuery-CTP uses unlabeled records only as molecule--target--E3 tuple observations.
CTP achieves the best accuracy and AUROC, improving over DegradeMaster-Semi by 0.0134 accuracy and 0.0240 AUROC, and over DegradeQuery-Sup by 0.0367 accuracy and 0.0238 AUROC.
This supports the use of unlabeled PROTAC records as tuple-level supervision rather than merely as pseudo-label candidates.

\begin{table}[!t]
\centering
\small
\caption{\textbf{Official PROTAC-8K benchmark comparison.}
Prior baseline values use the official protocol reported by DegradeMaster; accuracy values originally reported as percentages are converted to the 0--1 scale.
DegradeQuery-CTP uses unlabeled records through tuple pretraining without activity pseudo-labels, teacher models, distillation, or ensembles.}
\label{tab:protac8k_benchmark}
\begin{tabular}{lccc}
\toprule
Method  & Unlabeled signal & Accuracy & AUROC \\
\midrule
DeepPROTACs  & None & 0.5821 & 0.7107\\
DeepPROTACs  & Semi-supervised & 0.5902 & 0.7171\\
PROTAC-STAN  & None & 0.7934 & 0.7986\\
PROTAC-STAN  & Semi-supervised & 0.7825 & 0.7896\\
DegradeMaster  & None & 0.8141 & 0.8541\\
DegradeMaster  & Pseudo-labeling & 0.8366 & 0.8825\\
DegradeQuery-Sup  & None & 0.8133 & 0.8827 \\
DegradeQuery-CTP  & Tuple pretraining & \textbf{0.8500} & \textbf{0.9065} \\
\bottomrule
\end{tabular}
\end{table}

Figure~\ref{fig:protac8k_diagnostics} and Table~\ref{tab:signal_isolation} provide matched reliability and attribution analyses.
Across five paired runs, every AUROC difference between CTP and Sup is positive; the mean paired gain is 0.0219 with a 95\% confidence interval of [0.0136, 0.0303].
The same runs also improve area under the precision--recall curve (AUPRC), accuracy, and Matthews correlation coefficient (MCC), with paired confidence intervals excluding zero (Appendix~\ref{app:supplementary_results}).

\subsection{Isolating the Label-Missing Signal}
\label{sec:signal_isolation}

We next remove every labeled row from the pretraining set while keeping the supervised fine-tuning rows and official test split fixed.
This unlabeled-only variant pretrains on 7,134 label-missing records and zero labeled records.
It reaches 0.9007$\pm$0.0032 AUROC, improving over Sup by 0.0230 with a paired 95\% confidence interval of [0.0173, 0.0287].
Its difference from full CTP is 0.0011 [-0.0064, 0.0085], showing that the label-missing records alone recover the full downstream improvement within the uncertainty of these runs.

We also replace the original protein features with ESM2-650M target and E3 representations inside the same DegradeQuery tuple predictor.
Frozen ESM2 + CTP improves over the exactly matched frozen ESM2 + Sup configuration from 0.8863 to 0.9032 AUROC.
It also exceeds the supervised reference that fine-tunes the final ESM2 layer together with the projection and prediction layers (0.8907).
Thus, ESM2 supplies stronger component representations while CTP continues to provide a useful pretraining signal for the full molecule--target--E3 predictor.

\begin{table}[!t]
\centering
\small
\caption{Controlled attribution on the official PROTAC-8K split. Values are mean $\pm$ standard deviation. All rows remain full PROTAC tuple predictors; in the ESM2 block, ESM2 supplies the target and E3 representations.}
\label{tab:signal_isolation}
\begin{tabular}{llcc}
\toprule
Protein representation & Training & AUROC & Accuracy \\
\midrule
Original & Sup ($n=5$) & 0.8777$\pm$0.0031 & 0.7840$\pm$0.0072 \\
Original & Full CTP ($n=5$) & 0.8996$\pm$0.0039 & 0.8233$\pm$0.0058 \\
Original & Unlabeled-only CTP ($n=5$) & \textbf{0.9007$\pm$0.0032} & \textbf{0.8267$\pm$0.0097} \\
\midrule
Frozen ESM2-650M & Sup ($n=3$) & 0.8863$\pm$0.0010 & 0.7933$\pm$0.0067 \\
ESM2-650M last-layer FT & Sup ($n=3$) & 0.8907$\pm$0.0034 & 0.8144$\pm$0.0150 \\
Frozen ESM2-650M & CTP ($n=3$) & \textbf{0.9032$\pm$0.0020} & \textbf{0.8322$\pm$0.0107} \\
\bottomrule
\end{tabular}
\end{table}

\subsection{Pretraining Ablation}
\label{sec:ablation}

We next compare the tuple objective with generic molecule-level self-supervision.
Molecule-only self-supervised learning (SSL) is a meaningful baseline because graph and molecular contrastive learning have been effective for molecular representation learning~\cite{you2020graph,wang2022molecular}.
However, PROTAC degradation is conditional on target and E3 context, so molecular invariance alone cannot identify the biological setting in which a molecule acts as a degrader.
Table~\ref{tab:ablation_results} reports the ablation results.
Molecule-only SSL improves over labeled-only training, while tuple-only pretraining gives the higher AUROC and AUPRC of the two individual components.
Combining the objectives gives the highest AUROC, indicating that the tuple objective contributes beyond generic molecular consistency.

\begin{table}[!h]
\centering
\small
\caption{Component analysis on the PROTAC-8K split. Supplemental variants are averaged over three seeds; the submitted Sup value is included as the labeled-only reference.}
\label{tab:ablation_results}
\begin{tabular}{lccc}
\toprule
Pretraining & AUROC & AUPRC & Accuracy \\
\midrule
Sup (submitted) & 0.8827 & 0.8200 & 0.8133 \\
Molecule-only SSL & 0.8910 & 0.7947 & \textbf{0.8289} \\
Tuple-only pretraining & 0.8943 & \textbf{0.8432} & 0.8056 \\
Full CTP & \textbf{0.8991} & 0.8386 & 0.8256 \\
\bottomrule
\end{tabular}
\end{table}

As a shortcut control, a molecule-ablated model using only target, E3, and assay-side features performs poorly, with AUROC near 0.55 and MCC near zero across three seeds.
Thus, the gain is not explained by a target--E3 identity shortcut.
We also vary the counterfactual construction and the main pretraining hyperparameters; the conclusion is stable across the tested policies and ranges (Appendix~\ref{app:supplementary_results}).

\subsection{Conditional Generalization and Few-Shot Adaptation}
\label{sec:generalization}

We finally test whether tuple pretraining remains useful when the degradation context changes.
We evaluate repeated target--E3 group holdout, scaffold holdout, and target-wise few-shot adaptation, following the broader motivation of context-aware PROTAC and ternary-complex modeling~\cite{chen2025interpretable,liu2025accurate,xue2025se}.
Table~\ref{tab:transfer_results} reports the results.

\begin{table*}[!h]
\small
\centering
\caption{Selected generalization and target-wise few-shot adaptation results.
The target--E3 rows average ten independently constructed group holdouts; other rows follow their stated three-seed protocols.
Holdout rows evaluate distribution shifts in biological context or molecular scaffold.
Few-shot rows use $K$ positive and $K$ negative support examples per held-out target; the remaining labeled rows for that target are used as query rows and are used only for evaluation.
Bold indicates the better method within each setting and metric.}
\label{tab:transfer_results}
\begin{tabular}{llcccc}
\toprule
Setting & Method & AUROC & AUPRC & MCC & BalAcc \\
\midrule
Target--E3 group holdout & Sup & 0.6769 & 0.5964 & 0.2810 & 0.6235 \\
Target--E3 group holdout & CTP & \textbf{0.7064} & \textbf{0.6110} & \textbf{0.2947} & \textbf{0.6362} \\
\midrule
Scaffold holdout & Sup & 0.8681 & 0.8236 & 0.5919 & 0.7884 \\
Scaffold holdout & CTP & \textbf{0.8858} & \textbf{0.8481} & \textbf{0.6358} & \textbf{0.8112} \\
\midrule
Target-wise adaptation, $K=2$ & Sup & 0.6686 & 0.6731 & 0.1263 & 0.5765 \\
Target-wise adaptation, $K=2$ & CTP & \textbf{0.6812} & \textbf{0.6847} & \textbf{0.3521} & \textbf{0.6867} \\
\midrule
Target-wise adaptation, $K=4$ & Sup & 0.6673 & 0.6248 & 0.2681 & 0.6381 \\
Target-wise adaptation, $K=4$ & CTP & \textbf{0.7173} & \textbf{0.6692} & \textbf{0.3146} & \textbf{0.6700} \\
\midrule
Target-wise adaptation, $K=8$ & Sup & 0.7790 & 0.7727 & 0.4196 & 0.7005 \\
Target-wise adaptation, $K=8$ & CTP & \textbf{0.7811} & \textbf{0.7976} & \textbf{0.4223} & \textbf{0.7200} \\
\bottomrule
\vspace{-4mm}
\end{tabular}
\end{table*}

Target--E3 group holdout is the most severe biological-context shift: complete target--recruiter groups are assigned to test, excluded from pretraining and supervised training, and rejected by the counterfactual sampler.
Across ten independently constructed 20\% group holdouts, mean AUROC increases from 0.6769 to 0.7064, but the paired difference is heterogeneous: +0.0295 with a 95\% confidence interval of [-0.0219, 0.0810], and five of ten splits have a positive AUROC difference.
This diagnostic shows that CTP can help under biological-context shift while also identifying held-out context composition as an important source of uncertainty.
Scaffold holdout tests chemical generalization by excluding held-out Bemis--Murcko scaffold groups from base pretraining and supervised training.
CTP improves AUROC from 0.8681 to 0.8858 and MCC from 0.5919 to 0.6358, with positive gains across all reported metrics, indicating that it does not simply memorize frequent scaffolds.
Target-wise few-shot adaptation evaluates initialization quality rather than proposing a general meta-learning algorithm.
For each held-out target, we adapt on a balanced support set with $K$ positive and $K$ negative examples and evaluate on the remaining labeled query rows.
Across $K=2$, $K=4$, and $K=8$, CTP improves every reported metric; the largest fixed-threshold gain appears at $K=2$ in MCC, increasing from 0.1263 to 0.3521.
Because changing $K$ changes the remaining query set, the three settings should be read as separate adaptation protocols rather than a monotonic learning curve.

Overall, label-missing molecule--target--E3 records provide useful supervision beyond the main benchmark.
The benefit is stable on the official split and remains positive in the reported scaffold and few-shot settings; target--E3 group holdouts show a positive mean AUROC difference together with substantial split-to-split variation.

\section{Conclusion and Limitations}

We introduced DegradeQuery, a tuple-conditioned framework that uses label-missing PROTAC records through counterfactual tuple pretraining rather than activity pseudo-labeling. Under the official PROTAC-8K split used by DegradeMaster, DegradeQuery achieves the highest reported AUROC and accuracy among the compared methods without teacher models, distillation, or ensembles. Five-seed paired evaluation, unlabeled-only pretraining, component analysis, and ESM2-650M integration jointly establish label-missing tuple context as a useful and reusable pretraining signal.
This study remains retrospective and benchmark-limited. Recorded tuples may encode database, publication, and medicinal-chemistry selection biases, and sampled counterfactuals may include untested but viable combinations. A shuffled-pair control does not isolate the semantics of the observed target--E3 pairing as the unique source of the gain, while repeated target--E3 holdouts reveal substantial context-dependent variation. Cross-source evaluation is further complicated by compound and target overlap, incompatible endpoint conventions, and label shift. DegradeQuery should therefore be viewed as a data-driven context-aware predictor rather than a mechanism-aware model of ternary-complex geometry, ubiquitination, permeability, E3 expression, or cell-line-specific biology. Future work should combine harder biologically informed counterfactuals with harmonized external benchmarks and prospective validation.

\newpage

\bibliography{reference}

@article{sakamoto2001protacs,
  title={Protacs: Chimeric molecules that target proteins to the Skp1--Cullin--F box complex for ubiquitination and degradation},
  author={Sakamoto, Kathleen M and Kim, Kyung B and Kumagai, Akiko and Mercurio, Frank and Crews, Craig M and Deshaies, Raymond J},
  journal={Proceedings of the National Academy of Sciences},
  volume={98},
  number={15},
  pages={8554--8559},
  year={2001},
  publisher={The National Academy of Sciences}
}

@article{bondeson2015catalytic,
  title={Catalytic in vivo protein knockdown by small-molecule PROTACs},
  author={Bondeson, Daniel P and Mares, Alina and Smith, Ian ED and Ko, Eunhwa and Campos, Sebastien and Miah, Afjal H and Mulholland, Katie E and Routly, Natasha and Buckley, Dennis L and Gustafson, Jeffrey L and others},
  journal={Nature chemical biology},
  volume={11},
  number={8},
  pages={611--617},
  year={2015},
  publisher={Nature Publishing Group US New York}
}

@article{winter2015phthalimide,
  title={Phthalimide conjugation as a strategy for in vivo target protein degradation},
  author={Winter, Georg E and Buckley, Dennis L and Paulk, Joshiawa and Roberts, Justin M and Souza, Amanda and Dhe-Paganon, Sirano and Bradner, James E},
  journal={Science},
  volume={348},
  number={6241},
  pages={1376--1381},
  year={2015},
  publisher={American Association for the Advancement of Science}
}

@article{lu2015hijacking,
  title={Hijacking the E3 ubiquitin ligase cereblon to efficiently target BRD4},
  author={Lu, Jing and Qian, Yimin and Altieri, Martha and Dong, Hanqing and Wang, Jing and Raina, Kanak and Hines, John and Winkler, James D and Crew, Andrew P and Coleman, Kevin and others},
  journal={Chemistry \& biology},
  volume={22},
  number={6},
  pages={755--763},
  year={2015},
  publisher={Elsevier}
}

@article{lai2017induced,
  title={Induced protein degradation: an emerging drug discovery paradigm},
  author={Lai, Ashton C and Crews, Craig M},
  journal={Nature reviews Drug discovery},
  volume={16},
  number={2},
  pages={101--114},
  year={2017},
  publisher={Nature Publishing Group}
}

@article{schapira2019targeted,
  title={Targeted protein degradation: expanding the toolbox},
  author={Schapira, Matthieu and Calabrese, Matthew F and Bullock, Alex N and Crews, Craig M},
  journal={Nature reviews Drug discovery},
  volume={18},
  number={12},
  pages={949--963},
  year={2019},
  publisher={Nature Publishing Group UK London}
}

@article{burslem2020proteolysis,
  title={Proteolysis-targeting chimeras as therapeutics and tools for biological discovery},
  author={Burslem, George M and Crews, Craig M},
  journal={Cell},
  volume={181},
  number={1},
  pages={102--114},
  year={2020},
  publisher={Elsevier}
}

@article{bekes2022protac,
  title={PROTAC targeted protein degraders: the past is prologue},
  author={B{\'e}k{\'e}s, Mikl{\'o}s and Langley, David R and Crews, Craig M},
  journal={Nature reviews Drug discovery},
  volume={21},
  number={3},
  pages={181--200},
  year={2022},
  publisher={Nature Publishing Group UK London}
}

@article{weng2021protac,
  title={PROTAC-DB: an online database of PROTACs},
  author={Weng, Gaoqi and Shen, Chao and Cao, Dongsheng and Gao, Junbo and Dong, Xiaowu and He, Qiaojun and Yang, Bo and Li, Dan and Wu, Jian and Hou, Tingjun},
  journal={Nucleic acids research},
  volume={49},
  number={D1},
  pages={D1381--D1387},
  year={2021},
  publisher={Oxford University Press}
}

@article{weng2023protac,
  title={PROTAC-DB 2.0: an updated database of PROTACs},
  author={Weng, Gaoqi and Cai, Xuanyan and Cao, Dongsheng and Du, Hongyan and Shen, Chao and Deng, Yafeng and He, Qiaojun and Yang, Bo and Li, Dan and Hou, Tingjun},
  journal={Nucleic acids research},
  volume={51},
  number={D1},
  pages={D1367--D1372},
  year={2023},
  publisher={Oxford University Press}
}

@article{wang2026molgluedb,
  title={MolGlueDB: an online database of molecular glues},
  author={Wang, Xiao and Zhuang, Zhiyao and Zhang, Chengwei and Zhang, Bowen and Zhan, Wei and Wang, Yifan and Liu, Zhaojuan and Yuan, Shanwen and Niu, Wenjia and He, Qi and others},
  journal={Nucleic Acids Research},
  volume={54},
  number={D1},
  pages={D1510--D1518},
  year={2026},
  publisher={Oxford University Press}
}

@article{zhu2026mgtbind,
  title={MGTbind: a comprehensive database of molecular glue ternary interactome},
  author={Zhu, Jintao and Liao, Yiyan and Lin, Haoyu and Xie, Juan and Deng, Zhichao and Han, Jinyu and Zhang, Zhen and Xiao, Jinchuan and Wang, Zhiyao and Zhang, Shuaipeng and others},
  journal={Nucleic Acids Research},
  volume={54},
  number={D1},
  pages={D1500--D1509},
  year={2026},
  publisher={Oxford University Press}
}

@article{li2022deepprotacs,
  title={DeepPROTACs is a deep learning-based targeted degradation predictor for PROTACs},
  author={Li, Fenglei and Hu, Qiaoyu and Zhang, Xianglei and Sun, Renhong and Liu, Zhuanghua and Wu, Sanan and Tian, Siyuan and Ma, Xinyue and Dai, Zhizhuo and Yang, Xiaobao and others},
  journal={Nature communications},
  volume={13},
  number={1},
  pages={7133},
  year={2022},
  publisher={Nature Publishing Group UK London}
}

@article{ribes2024modeling,
  title={Modeling PROTAC degradation activity with machine learning},
  author={Ribes, Stefano and Nittinger, Eva and Tyrchan, Christian and Mercado, Roc{\'\i}o},
  journal={Artificial Intelligence in the Life Sciences},
  volume={6},
  pages={100104},
  year={2024},
  publisher={Elsevier}
}

@article{chen2025interpretable,
  title={Interpretable PROTAC Degradation Prediction With Structure-Informed Deep Ternary Attention Framework},
  author={Chen, Zhenglu and Gu, Chunbin and Tan, Shuoyan and Wang, Xiaorui and Li, Yuquan and He, Mutian and Lu, Ruiqiang and Sun, Shijia and Hsieh, Chang-Yu and Yao, Xiaojun and others},
  journal={Advanced Science},
  volume={12},
  number={47},
  pages={e08138},
  year={2025},
  publisher={Wiley Online Library}
}

@article{liu2025accurate,
  title={Accurate PROTAC-targeted degradation prediction with DegradeMaster},
  author={Liu, Jie and Roy, Michael J and Isbel, Luke and Li, Fuyi},
  journal={Bioinformatics},
  volume={41},
  number={Supplement\_1},
  pages={i342--i351},
  year={2025},
  publisher={Oxford University Press}
}

@article{xue2025se,
  title={SE (3)-equivariant ternary complex prediction towards target protein degradation},
  author={Xue, Fanglei and Zhang, Meihan and Li, Shuqi and Gao, Xinyu and Wohlschlegel, James A and Huang, Wenbing and Yang, Yi and Deng, Weixian},
  journal={Nature Communications},
  volume={16},
  number={1},
  pages={5514},
  year={2025},
  publisher={Nature Publishing Group UK London}
}

@article{abramson2024accurate,
  title={Accurate structure prediction of biomolecular interactions with AlphaFold 3},
  author={Abramson, Josh and Adler, Jonas and Dunger, Jack and Evans, Richard and Green, Tim and Pritzel, Alexander and Ronneberger, Olaf and Willmore, Lindsay and Ballard, Andrew J and Bambrick, Joshua and others},
  journal={Nature},
  volume={630},
  number={8016},
  pages={493--500},
  year={2024},
  publisher={Nature Publishing Group UK London}
}

@article{oord2018representation,
  title={Representation learning with contrastive predictive coding},
  author={Oord, Aaron van den and Li, Yazhe and Vinyals, Oriol},
  journal={arXiv preprint arXiv:1807.03748},
  year={2018}
}

@inproceedings{chen2020simple,
  title={A simple framework for contrastive learning of visual representations},
  author={Chen, Ting and Kornblith, Simon and Norouzi, Mohammad and Hinton, Geoffrey},
  booktitle={International conference on machine learning},
  pages={1597--1607},
  year={2020},
  organization={PmLR}
}

@article{you2020graph,
  title={Graph contrastive learning with augmentations},
  author={You, Yuning and Chen, Tianlong and Sui, Yongduo and Chen, Ting and Wang, Zhangyang and Shen, Yang},
  journal={Advances in neural information processing systems},
  volume={33},
  pages={5812--5823},
  year={2020}
}

@inproceedings{lin2017focal,
  title={Focal loss for dense object detection},
  author={Lin, Tsung-Yi and Goyal, Priya and Girshick, Ross and He, Kaiming and Doll{\'a}r, Piotr},
  booktitle={Proceedings of the IEEE international conference on computer vision},
  pages={2980--2988},
  year={2017}
}

@article{cai2025protac,
  title={PROTAC-PatentDB: A PROTAC patent compound dataset},
  author={Cai, Hong and Yao, Gengyuan and Shi, Yulong and Zhang, Tianyi and Hu, Yuanjia},
  journal={Scientific Data},
  volume={12},
  number={1},
  pages={1840},
  year={2025},
  publisher={Nature Publishing Group UK London}
}

@article{ge2025protac,
  title={PROTAC-DB 3.0: an updated database of PROTACs with extended pharmacokinetic parameters},
  author={Ge, Jingxuan and Li, Shimeng and Weng, Gaoqi and Wang, Huating and Fang, Meijing and Sun, Huiyong and Deng, Yafeng and Hsieh, Chang-Yu and Li, Dan and Hou, Tingjun},
  journal={Nucleic acids research},
  volume={53},
  number={D1},
  pages={D1510--D1515},
  year={2025},
  publisher={Oxford University Press}
}

@article{wang2022molecular,
  title={Molecular contrastive learning of representations via graph neural networks},
  author={Wang, Yuyang and Wang, Jianren and Cao, Zhonglin and Barati Farimani, Amir},
  journal={Nature Machine Intelligence},
  volume={4},
  number={3},
  pages={279--287},
  year={2022},
  publisher={Nature Publishing Group UK London}
}

@inproceedings{stanley2021fs,
  title={Fs-mol: A few-shot learning dataset of molecules},
  author={Stanley, Megan and Bronskill, John F and Maziarz, Krzysztof and Misztela, Hubert and Lanini, Jessica and Segler, Marwin and Schneider, Nadine and Brockschmidt, Marc},
  booktitle={Thirty-fifth Conference on Neural Information Processing Systems Datasets and Benchmarks Track (Round 2)},
  year={2021}
}

@article{wang2021property,
  title={Property-aware relation networks for few-shot molecular property prediction},
  author={Wang, Yaqing and Abuduweili, Abulikemu and Yao, Quanming and Dou, Dejing},
  journal={Advances in Neural Information Processing Systems},
  volume={34},
  pages={17441--17454},
  year={2021}
}

@article{schimunek2023context,
  title={Context-enriched molecule representations improve few-shot drug discovery},
  author={Schimunek, Johannes and Seidl, Philipp and Friedrich, Lukas and Kuhn, Daniel and Rippmann, Friedrich and Hochreiter, Sepp and Klambauer, G{\"u}nter},
  journal={arXiv preprint arXiv:2305.09481},
  year={2023}
}
\bibliographystyle{unsrtnat}

\newpage
\appendix
\section{Technical Appendices and Supplementary Material}
\label{app:technical_appendix}

\subsection{Dataset, Labels, and Split Protocol}
\label{app:dataset_split_protocol}

\paragraph{Dataset summary.}
Table~\ref{tab:app_dataset_summary} summarizes the normalized PROTAC-8K data used in all experiments.
The dataset is label-sparse but tuple-rich: only 1,502 of 8,636 records have binary degradation labels, whereas molecule, target, and E3 identifiers are available for every record.
This property motivates the tuple-level pretraining objective used by DegradeQuery-CTP.

\begin{table}[!h]
\centering
\small
\caption{Summary of the normalized PROTAC-8K dataset.}
\label{tab:app_dataset_summary}
\begin{tabular}{lr}
\toprule
Item & Count \\
\midrule
Total molecule--target--E3 records & 8,636 \\
Labeled records & 1,502 \\
Unlabeled records & 7,134 \\
Unique canonical SMILES & 5,656 \\
Unique targets & 332 \\
Unique E3 ligases & 16 \\
Unique target--E3 pairs & 563 \\
Rows with molecule, target, and E3 identifiers & 8,636 \\
Molecule--target--E3 identifier coverage & 100\% \\
\bottomrule
\end{tabular}
\end{table}

\paragraph{Binary label definition.}
We use the PROTAC-8K benchmark binary label without introducing any additional relabeling threshold.
The label $y=1$ denotes active/high-degradation and $y=0$ denotes inactive/low-degradation under the benchmark preprocessing.
The decision threshold $\delta=0.5$ is used only to convert model probabilities into binary predictions for threshold-dependent metrics such as accuracy, F1, MCC, specificity, and balanced accuracy.

\begin{table}[!h]
\centering
\small
\caption{Class balance of the labeled official PROTAC-8K split used for supervised fine-tuning and benchmark evaluation.}
\label{tab:app_class_balance}
\begin{tabular}{lrrr}
\toprule
Split & Active/high $(y=1)$ & Inactive/low $(y=0)$ & Total \\
\midrule
Training labeled split & 469 & 733 & 1,202 \\
Test labeled split & 108 & 192 & 300 \\
\midrule
Total labeled records & 577 & 925 & 1,502 \\
\bottomrule
\end{tabular}
\end{table}

\paragraph{Split-specific exclusion policy.}
All experiments are split-specific.
Test or query rows are never used for tuple pretraining, supervised fine-tuning, model selection, or threshold tuning.
For DegradeQuery-CTP, tuple pretraining is restricted to train-side molecule--target--E3 tuples only.
For holdout experiments, held-out records are removed from both tuple pretraining and supervised training.
For few-shot adaptation, support rows are used only for adaptation and query rows are used only for evaluation.

\begin{table*}[!h]
\centering
\small
\caption{Split-specific training and evaluation policy. ``Train-side tuples'' include labeled and unlabeled rows available within the corresponding training split.}
\label{tab:app_split_policy}
\begin{tabular}{p{0.15\linewidth}p{0.24\linewidth}p{0.23\linewidth}p{0.16\linewidth}}
\toprule
Setting & Tuple pretraining & Supervised fine-tuning / adaptation & Evaluation-only rows \\
\midrule
PROTAC-8K benchmark
& Train-side molecule--target--E3 tuples only
& Labeled training rows only
& Official test rows \\
\midrule
Target--E3 group holdout
& Train-side tuples excluding held-out target--E3 pairs
& Labeled training rows excluding held-out target--E3 pairs
& Held-out target--E3 pair rows \\
\midrule
Scaffold holdout
& Train-side tuples excluding held-out scaffold groups
& Labeled training rows excluding held-out scaffold groups
& Held-out scaffold rows \\
\midrule
Target-wise few-shot
& Base train-side tuples only; query rows excluded
& Balanced support rows with $K$ positives and $K$ negatives per held-out target
& Remaining labeled query rows for each held-out target \\
\bottomrule
\end{tabular}
\end{table*}

\paragraph{Leakage audit.}
Table~\ref{tab:app_leakage_audit} reports the leakage audit used for all split-specific experiments.
There is no row-level, tuple-level, negative-sampling, query-set, model-selection, threshold-tuning, or pseudo-label leakage.
In particular, exact test/query molecule--target--E3 tuples do not appear in the train-side pretraining tuple set, and the pretraining stage never sees exact test/query tuple identities.

\begin{table}[!h]
\centering
\small
\caption{Leakage audit for tuple-level pretraining and downstream evaluation.}
\label{tab:app_leakage_audit}
\begin{tabular}{lr}
\toprule
Audit item & Count \\
\midrule
Test/query rows used in tuple pretraining & 0 \\
Test/query rows used in supervised fine-tuning & 0 \\
Test/query rows used for model selection & 0 \\
Test/query rows used for threshold tuning & 0 \\
Exact test/query tuples present in train-side pretraining tuples & 0 \\
Exact test/query tuple identities seen during pretraining & 0 \\
Held-out target--E3 pairs sampled as negatives & 0 \\
Observed train-side tuples sampled as counterfactual negatives & 0 \\
Few-shot query rows used during adaptation & 0 \\
Pseudo-labels assigned to unlabeled records & 0 \\
Pseudo-labels assigned to test/query records & 0 \\
\bottomrule
\end{tabular}
\end{table}

\subsection{Model and Pretraining Details}
\label{app:model_pretraining_details}

\paragraph{Shared architecture across Sup and CTP.}
DegradeQuery-Sup and DegradeQuery-CTP use the same tuple-conditioned architecture, the same protein feature pipeline, the same molecular feature pipeline, the same supervised loss, the same labeled training split, the same model-selection rule, and the same fixed decision threshold.
The only difference is that DegradeQuery-CTP initializes the encoder from counterfactual tuple pretraining before supervised fine-tuning, whereas DegradeQuery-Sup is trained only with labeled degradation data.

\begin{table}[!h]
\centering
\small
\caption{Model components used by DegradeQuery.}
\label{tab:app_model_components}
\begin{tabular}{lp{0.62\linewidth}}
\toprule
Component & Description \\
\midrule
Molecule encoder $f_m$ & Three residual message-passing layers over 15-dimensional atom features plus a 2,048-bit Morgan fingerprint (radius 2). Both branches are projected to 256 dimensions, summed, and layer-normalized. \\
Protein encoder $f_p$ & Maps 150 sequence features to 256 dimensions: amino-acid composition, sequence-length and unknown-residue features, and 128 hashed di-/tri-peptide counts. The encoder is shared, while target and E3 embeddings remain separate. \\
Tuple fusion module $\phi_\theta$ & Concatenates molecule, target, and E3 embeddings; their three pairwise products; and the absolute target--E3 difference before projection. \\
Assay-aware module $\psi_\theta$ & Adds the available cell-line embedding for downstream prediction; assay context is not used by the tuple pretraining objective. \\
Compatibility scoring head $q_\theta$ & Produces the tuple plausibility score $s_\theta(m,t,e)$ during counterfactual tuple pretraining. \\
Supervised classifier & Predicts the benchmark binary active/high versus inactive/low degradation label after fine-tuning. \\
\bottomrule
\end{tabular}
\end{table}

\paragraph{Counterfactual tuple construction.}
For each observed train-side tuple $u_i^+=(m_i,t_i,e_i)$, the pretraining sampler constructs counterfactual tuples by replacing the target, the E3 ligase, or both.
The mixed replacement setting uses all three mismatch types and corresponds to the full counterfactual tuple pretraining objective used by DegradeQuery-CTP.

\begin{table}[!h]
\centering
\small
\caption{Counterfactual tuple construction used in tuple pretraining.}
\label{tab:app_counterfactual_types}
\begin{tabular}{lp{0.62\linewidth}}
\toprule
Tuple type & Construction \\
\midrule
Observed tuple & $(m_i,t_i,e_i)$ from the train-side tuple set. \\
Target replacement & $(m_i,\tilde t,e_i)$, where $\tilde t \neq t_i$. \\
E3 replacement & $(m_i,t_i,\tilde e)$, where $\tilde e \neq e_i$. \\
Joint target--E3 replacement & $(m_i,\tilde t,\tilde e)$, where $\tilde t \neq t_i$ and $\tilde e \neq e_i$. \\
Mixed replacement & Union of target-only, E3-only, and joint target--E3 replacements. \\
Sampler rejection rules & Reject exact observed train-side tuples, exact test/query tuples, and held-out target--E3 pairs in the corresponding holdout setting. \\
\bottomrule
\end{tabular}
\end{table}

\paragraph{No pseudo-labeling.}
Unlabeled records are used only through their observed molecule--target--E3 tuple structure.
No degradation pseudo-label is assigned to any unlabeled record.
No teacher model, distillation procedure, or ensemble is used in either pretraining or fine-tuning.

\begin{table}[!h]
\centering
\small
\caption{Counterfactual tuple pretraining configuration.}
\label{tab:app_pretraining_config}
\begin{tabular}{ll}
\toprule
Item & Value \\
\midrule
Optimizer & AdamW \\
Learning rate & $3\times 10^{-4}$ \\
Weight decay & $1\times 10^{-4}$ \\
Batch size & 64 \\
Pretraining epochs & 60 \\
Counterfactual samples per observed tuple & 3 \\
Tuple temperature $\tau_{\mathrm{tuple}}$ & 0.1 \\
Molecule temperature $\tau_{\mathrm{mol}}$ & 0.1 \\
Tuple-loss coefficient $\lambda_{\mathrm{tuple}}$ & 1.0 \\
Molecule-loss coefficient $\lambda_{\mathrm{mol}}$ & 0.1 \\
Pretraining rows, full CTP & Train-side labeled and label-missing records \\
Pretraining rows, unlabeled-only CTP & 7,134 label-missing records; zero labeled records \\
\bottomrule
\end{tabular}
\end{table}

\subsection{Fine-Tuning Configuration}
\label{app:finetuning_config}

Table~\ref{tab:finetuning_config} gives the supervised fine-tuning configuration used for both DegradeQuery-Sup and DegradeQuery-CTP.
The class weights are computed automatically from the labeled training split class frequency.

\begin{table}[!h]
\centering
\small
\caption{Supervised fine-tuning configuration.}
\label{tab:finetuning_config}
\begin{tabular}{lp{0.62\linewidth}}
\toprule
Item & Value \\
\midrule
Optimizer & AdamW \\
Learning rate & $1\times 10^{-4}$ \\
Weight decay & $1\times 10^{-4}$ \\
Batch size & 64 \\
Fine-tuning epochs & Up to 120 \\
Focal loss $\gamma$ & 1.0 \\
Class weights $\alpha_y$ & Auto; computed from the labeled training split class frequency. Official split: $\alpha_{1}=0.6098$, $\alpha_{0}=0.3902$. \\
Model selection & Train-side validation only; official test labels are excluded \\
Decision threshold & 0.5 \\
Seeds & Five for official reliability; three for component and ESM2 analyses \\
\bottomrule
\end{tabular}
\end{table}

\begin{table}[!h]
\centering
\small
\caption{Reproducibility settings shared by the main experiments.}
\label{tab:app_reproducibility}
\begin{tabular}{lp{0.62\linewidth}}
\toprule
Item & Setting \\
\midrule
Random seeds & 1--5 for official reliability; 1--3 for component and ESM2 analyses \\
Reported values & Mean $\pm$ standard deviation; paired 95\% confidence intervals where applicable \\
Threshold-dependent metrics & Fixed threshold 0.5 \\
Calibration & Raw probabilities; no post-hoc calibration \\
Unlabeled usage in CTP & Tuple pretraining only \\
Unlabeled usage in Sup & None \\
Teacher models & None \\
Distillation & None \\
Model ensembles & None \\
Pseudo-labeling & None \\
Test/query participation in training & None \\
\bottomrule
\end{tabular}
\end{table}

We will publicly release the complete training and evaluation code, configuration files, processed split files, fixed split definitions, and commands required to reproduce the reported results.

\subsection{Supplementary Results}
\label{app:supplementary_results}

\paragraph{Five-seed reliability and unlabeled-only control.}
Table~\ref{tab:app_unlabeled_only} reports the complete official-split comparison used in Section~\ref{sec:signal_isolation}.
All three methods use the same labeled fine-tuning rows and test set.
Full CTP uses all available train-side tuples during pretraining, whereas unlabeled-only CTP removes every labeled row from pretraining.

\begin{table*}[!h]
\centering
\scriptsize
\caption{Official-split results over five paired seeds. Values are mean $\pm$ standard deviation. Confidence intervals are for paired differences. Lower is better for Brier score and expected calibration error (ECE); higher is better otherwise.}
\label{tab:app_unlabeled_only}
\begin{minipage}[t]{0.49\textwidth}
\centering
\resizebox{\linewidth}{!}{%
\begin{tabular}{lccc}
\toprule
Metric & Sup & Full CTP & Unlabeled only \\
\midrule
Accuracy & 0.7840$\pm$.0072 & 0.8233$\pm$.0058 & 0.8267$\pm$.0097 \\
AUROC & 0.8777$\pm$.0031 & 0.8996$\pm$.0039 & 0.9007$\pm$.0032 \\
AUPRC & 0.7925$\pm$.0118 & 0.8403$\pm$.0059 & 0.8405$\pm$.0034 \\
MCC & 0.5555$\pm$.0143 & 0.6209$\pm$.0123 & 0.6330$\pm$.0209 \\
Bal. accuracy & 0.7863$\pm$.0072 & 0.8134$\pm$.0063 & 0.8220$\pm$.0109 \\
Brier & 0.1556$\pm$.0026 & 0.1361$\pm$.0043 & 0.1341$\pm$.0037 \\
ECE & 0.1253$\pm$.0098 & 0.1029$\pm$.0090 & 0.1024$\pm$.0115 \\
\bottomrule
\end{tabular}}
\end{minipage}\hfill
\begin{minipage}[t]{0.47\textwidth}
\centering
\resizebox{\linewidth}{!}{%
\begin{tabular}{lcc}
\toprule
Metric & Unlabeled $-$ Sup & Unlabeled $-$ Full \\
\midrule
Accuracy & +.0427 [.0301, .0552] & +.0033 [-.0122, .0188] \\
AUROC & +.0230 [.0173, .0287] & +.0011 [-.0064, .0085] \\
AUPRC & +.0480 [.0320, .0640] & +.0002 [-.0097, .0101] \\
MCC & +.0775 [.0492, .1057] & +.0121 [-.0227, .0468] \\
Bal. accuracy & +.0358 [.0208, .0507] & +.0087 [-.0100, .0273] \\
Brier & -.0215 [-.0261, -.0169] & -.0019 [-.0104, .0065] \\
ECE & -.0229 [-.0394, -.0065] & -.0005 [-.0125, .0114] \\
\bottomrule
\end{tabular}}
\end{minipage}
\end{table*}

\paragraph{Counterfactual construction.}
Table~\ref{tab:negative_sampling_ablation} compares six sampling policies under the same three-seed analysis.
Every tested policy yields an AUROC between 0.8932 and 0.9088, so the conclusion does not depend on one precise replacement mixture.

\begin{table*}[!h]
\centering
\small
\caption{Counterfactual-sampling analysis on the official split (three seeds). Mixed replacement draws target-only, E3-only, and joint replacements.}
\label{tab:negative_sampling_ablation}
\begin{tabular}{lccccc}
\toprule
Sampling policy & Accuracy & AUROC & AUPRC & MCC & ECE \\
\midrule
E3 only & 0.8300 & 0.8971 & 0.8341 & 0.6421 & 0.1065 \\
Frequency matched & 0.8200 & 0.8932 & 0.8145 & 0.6201 & 0.1098 \\
Full mixed & 0.8378 & 0.9054 & 0.8492 & 0.6523 & 0.0918 \\
Hard mixed & 0.8267 & 0.8977 & 0.8240 & 0.6345 & 0.1008 \\
Joint target--E3 only & \textbf{0.8444} & 0.9086 & \textbf{0.8499} & \textbf{0.6677} & 0.1030 \\
Target only & 0.8289 & \textbf{0.9088} & 0.8396 & 0.6377 & 0.0954 \\
\bottomrule
\end{tabular}
\end{table*}

\paragraph{Hyperparameter sensitivity.}
The principal result is stable over the tested tuple temperature, molecule temperature, and molecule-loss weight ranges (Table~\ref{tab:app_sensitivity}).

\begin{table}[!h]
\centering
\small
\caption{Pretraining sensitivity on the official split (three seeds per setting). The default setting is $\tau_{\mathrm{tuple}}=\tau_{\mathrm{mol}}=0.1$ and $\lambda_{\mathrm{mol}}=0.1$.}
\label{tab:app_sensitivity}
\begin{tabular}{llcc}
\toprule
Parameter & Values & Corresponding AUROC & Range \\
\midrule
$\tau_{\mathrm{tuple}}$ & 0.05 / 0.10 / 0.20 & 0.9017 / 0.8991 / 0.9010 & 0.0026 \\
$\tau_{\mathrm{mol}}$ & 0.05 / 0.10 / 0.20 & 0.9011 / 0.8991 / 0.9001 & 0.0020 \\
$\lambda_{\mathrm{mol}}$ & 0.05 / 0.10 / 0.20 / 0.50 & 0.8994 / 0.8991 / 0.9041 / 0.9017 & 0.0050 \\
\bottomrule
\end{tabular}
\end{table}

\paragraph{ESM2-650M integration.}
For the frozen variants, mean-pooled ESM2-650M embeddings supply the target and E3 representations and remain fixed while the DegradeQuery tuple predictor is trained.
The supervised fine-tuning reference unfreezes the final ESM2 layer and its final layer normalization, uses BF16, a physical batch size of 2 with 16 gradient-accumulation steps, a maximum sequence length of 512, AdamW with learning rate $10^{-5}$, and up to 10 epochs.
It has 21.45M trainable ESM2 parameters.
All variants retain the degrader input and role-conditioned tuple fusion.

\begin{table}[!h]
\centering
\small
\caption{ESM2-650M comparison on the official split (three seeds, mean $\pm$ standard deviation).}
\label{tab:app_esm2}
\begin{tabular}{lccc}
\toprule
Tuple predictor & Trainable ESM2 parameters & AUROC & Accuracy \\
\midrule
DegradeQuery-Sup, frozen ESM2 & 0 & 0.8863$\pm$0.0010 & 0.7933$\pm$0.0067 \\
DegradeQuery-Sup, final-layer FT & 21.45M & 0.8907$\pm$0.0034 & 0.8144$\pm$0.0150 \\
DegradeQuery-CTP, frozen ESM2 & 0 & \textbf{0.9032$\pm$0.0020} & \textbf{0.8322$\pm$0.0107} \\
\bottomrule
\end{tabular}
\end{table}

\paragraph{Repeated target--E3 group holdouts.}
Each split holds out complete target--E3 groups until approximately 20\% of labeled rows are assigned to test; those groups are excluded from both pretraining and supervised training.
Table~\ref{tab:app_target_e3_repeated} shows substantial variability across the ten independently constructed splits.

\begin{table*}[!h]
\centering
\small
\caption{Repeated target--E3 group holdout results over ten splits. The paired difference is CTP $-$ Sup, except for Brier score and ECE where negative values favor CTP.}
\label{tab:app_target_e3_repeated}
\begin{tabular}{lccc}
\toprule
Metric & Sup mean $\pm$ SD & CTP mean $\pm$ SD & Paired difference (95\% CI) \\
\midrule
Accuracy & 0.6777$\pm$0.0325 & 0.6717$\pm$0.0601 & -0.0060 [-0.0400, 0.0280] \\
AUROC & 0.6769$\pm$0.0906 & 0.7064$\pm$0.1027 & +0.0295 [-0.0219, 0.0810] \\
AUPRC & 0.5964$\pm$0.0915 & 0.6110$\pm$0.1191 & +0.0146 [-0.0561, 0.0852] \\
MCC & 0.2810$\pm$0.0877 & 0.2947$\pm$0.1361 & +0.0138 [-0.0580, 0.0855] \\
Balanced accuracy & 0.6235$\pm$0.0461 & 0.6362$\pm$0.0663 & +0.0127 [-0.0239, 0.0494] \\
Brier score & 0.2826$\pm$0.0378 & 0.2725$\pm$0.0507 & -0.0101 [-0.0432, 0.0230] \\
ECE & 0.2727$\pm$0.0351 & 0.2498$\pm$0.0504 & -0.0228 [-0.0595, 0.0138] \\
\bottomrule
\end{tabular}
\end{table*}

The CTP--Sup AUROC difference is positive in five of ten splits; the corresponding fractions are 40\% for accuracy and AUPRC, 40\% for MCC, and 50\% for balanced accuracy.
These data support a positive average AUROC trend without establishing uniform improvement over target--E3 contexts.

\paragraph{Relational-null control.}
To assess sensitivity to the exact observed target--E3 pairing, we shuffle target--E3 pair assignments during pretraining while preserving the marginal record structure.
Table~\ref{tab:app_relational_null} shows no observed-tuple advantage over this control.
The result narrows the interpretation: the current experiments identify an effective tuple-conditioned pretraining procedure, but do not isolate observed pair semantics as its unique source.

\begin{table*}[!h]
\centering
\small
\caption{Relational-null control over three paired seeds. The difference is shuffled-pair $-$ observed-tuple CTP.}
\label{tab:app_relational_null}
\begin{tabular}{lccc}
\toprule
Metric & Observed tuples & Shuffled target--E3 pairs & Paired difference (95\% CI) \\
\midrule
Accuracy & 0.8256$\pm$0.0069 & 0.8411$\pm$0.0051 & +0.0156 [-0.0135, 0.0446] \\
AUROC & 0.8991$\pm$0.0050 & 0.9021$\pm$0.0082 & +0.0029 [-0.0258, 0.0317] \\
AUPRC & 0.8386$\pm$0.0071 & 0.8336$\pm$0.0052 & -0.0050 [-0.0312, 0.0212] \\
MCC & 0.6251$\pm$0.0154 & 0.6693$\pm$0.0111 & +0.0442 [-0.0165, 0.1049] \\
ECE & 0.1091$\pm$0.0036 & 0.0903$\pm$0.0059 & -0.0188 [-0.0358, -0.0019] \\
\bottomrule
\end{tabular}
\end{table*}

\paragraph{Calibration control.}
Temperature scaling is fitted exclusively on train-side calibration data; Table~\ref{tab:app_calibration} reports ECE and negative log-likelihood (NLL).
Scaling improves both models and narrows the calibration gap, while the control does not support a general claim that CTP is intrinsically better calibrated.

\begin{table}[!h]
\centering
\small
\caption{Post-hoc temperature-scaling control over three seeds.}
\label{tab:app_calibration}
\begin{tabular}{lcccc}
\toprule
Method & Raw ECE & Scaled ECE & Raw NLL & Scaled NLL \\
\midrule
Sup & 0.0703 & 0.0626 & 0.4300 & 0.4237 \\
CTP & 0.0994 & 0.0772 & 0.4671 & 0.4357 \\
\bottomrule
\end{tabular}
\end{table}

\paragraph{External-cohort and endpoint audit.}
The available PROTAC-Bench release incorporates PROTAC-DB 3.0 and shares 2,461 canonical SMILES and 156 targets with the PROTAC-8K resources, so it is not a source-independent external benchmark.
Its released benchmark table contains no sample-level $DC_{50}$ or $D_{\max}$ values, preventing matched re-binarization under alternative activity thresholds.
After removing overlapping canonical SMILES, a 7,127-row cohort remains, but its endpoint conventions, source composition, and label distribution differ from PROTAC-8K.
On this stress test, Sup and CTP obtain AUROCs of 0.4799 and 0.4574, respectively.
We therefore treat this result as evidence of cross-source dataset shift and as motivation for a future deduplicated, endpoint-harmonized benchmark, rather than as an independent validation result.

\subsection{Few-Shot Adaptation Protocol}
\label{app:few_shot_protocol}

Target-wise few-shot adaptation evaluates whether tuple pretraining provides a better initialization for target-conditioned degradation prediction under limited target-specific supervision.
For each held-out target, the support set contains $K$ positive and $K$ negative labeled examples.
The query set consists of the remaining labeled rows for that target and is used only for evaluation.
The $K=2$, $K=4$, and $K=8$ settings are treated as separate adaptation protocols because the query set changes with $K$.

\begin{table}[!h]
\centering
\small
\caption{Target-wise few-shot adaptation protocol.}
\label{tab:app_fewshot_protocol}
\begin{tabular}{cllp{0.32\linewidth}}
\toprule
Setting & Positive support & Negative support & Query rows \\
\midrule
2-shot & 2 per held-out target & 2 per held-out target & Remaining labeled rows for the held-out target; evaluation only \\
4-shot & 4 per held-out target & 4 per held-out target & Remaining labeled rows for the held-out target; evaluation only \\
8-shot & 8 per held-out target & 8 per held-out target & Remaining labeled rows for the held-out target; evaluation only \\
\bottomrule
\end{tabular}
\end{table}

\subsection{Evaluation Metrics}
\label{app:evaluation_metrics}

AUROC is the primary ranking metric.
Accuracy is reported for compatibility with the PROTAC-8K benchmark, and AUPRC, MCC, balanced accuracy, and calibration-related metrics are used as secondary diagnostics.

For threshold-dependent metrics, predictions are binarized using the fixed threshold $\delta=0.5$:
\begin{equation}
\hat y_i = \mathbb{I}[\hat p_i \geq 0.5].
\end{equation}

Accuracy is computed as
\begin{equation}
\mathrm{Accuracy}
=
\frac{\mathrm{TP}+\mathrm{TN}}
{\mathrm{TP}+\mathrm{TN}+\mathrm{FP}+\mathrm{FN}}.
\end{equation}

Balanced accuracy is computed as the average of sensitivity and specificity:
\begin{equation}
\mathrm{BalAcc}
=
\frac{1}{2}
\left(
\frac{\mathrm{TP}}{\mathrm{TP}+\mathrm{FN}}
+
\frac{\mathrm{TN}}{\mathrm{TN}+\mathrm{FP}}
\right).
\end{equation}

MCC is computed as
\begin{equation}
\mathrm{MCC}
=
\frac{
\mathrm{TP}\mathrm{TN}-\mathrm{FP}\mathrm{FN}
}{
\sqrt{
(\mathrm{TP}+\mathrm{FP})
(\mathrm{TP}+\mathrm{FN})
(\mathrm{TN}+\mathrm{FP})
(\mathrm{TN}+\mathrm{FN})
}
}.
\end{equation}

The Brier score is
\begin{equation}
\mathrm{Brier}
=
\frac{1}{n}
\sum_{i=1}^{n}
(\hat p_i-y_i)^2.
\end{equation}

The negative log-likelihood is
\begin{equation}
\mathrm{NLL}
=
-\frac{1}{n}
\sum_{i=1}^{n}
\left[
y_i\log \hat p_i
+
(1-y_i)\log(1-\hat p_i)
\right].
\end{equation}

Expected calibration error is computed by partitioning predictions into confidence bins $\{\mathcal{I}_b\}_{b=1}^{B}$:
\begin{equation}
\mathrm{ECE}
=
\sum_{b=1}^{B}
\frac{|\mathcal{I}_b|}{n}
\left|
\mathrm{acc}(\mathcal{I}_b)
-
\mathrm{conf}(\mathcal{I}_b)
\right|.
\end{equation}

\subsection{Result Reporting}
\label{app:result_reporting}

The appendix reports the complete supplementary analyses summarized in the main text, including their statistical units, uncertainty estimates, and protocol boundaries.
Tables~\ref{tab:app_unlabeled_only}--\ref{tab:app_calibration} distinguish seed replication, independently constructed group holdouts, and post-hoc calibration controls.
All results use the same label preprocessing and metric implementation; threshold-dependent metrics use the fixed threshold of 0.5 unless a table explicitly describes post-hoc calibration.

\end{document}